\documentstyle[epsf,11pt,aaspp4]{article}

\begin{document}
\title{{\it RXTE} Observation of Cygnus X-1 In Its High State}
\author{W.~Cui\altaffilmark{1}, W.~A.~Heindl\altaffilmark{2}, R.~E.~Rothschild\altaffilmark{2}, S.~N.~Zhang\altaffilmark{3}, K.~Jahoda\altaffilmark{4}, and W.~Focke\altaffilmark{4}\altaffilmark{5}}
\altaffiltext{1}{Center for Space Research, Massachusetts Institute of Technology, Cambridge, MA 02139}
\altaffiltext{2}{Center for Astrophysics and Space Sciences, University of California, San Diego, La Jolla, CA 92093}
\altaffiltext{3}{NASA/Marshall Space Flight Center, Huntsville, AL 35812}
\altaffiltext{4}{NASA/Goddard Space Flight Center, Greenbelt, MD 20771}
\altaffiltext{5}{also Department of Physics, University of Maryland, College Park, MD 20741}
\authoremail{cui@space.mit.edu}
\slugcomment{revised and submitted to the {\it Astrophysical Journal Letters} on 9/26/96.}
\begin{abstract}

We present the results from the {\it RXTE} observations of Cygnus X-1 in its 
high state. In the energy range of 2-200 keV, the observed X-ray spectrum can 
be described by a model consisting of a 
soft blackbody component and a broken power-law with a high energy cutoff. 
The low energy spectrum (below about 11 keV) varies significantly from 
observation to observation while the high energy portion changes little. 
The X-ray flux varies on all timescales down to milliseconds. The power 
density spectrum (PDS) can be characterized by excess red noise (``1/f'') at 
low frequencies and a white noise component that extends to 1-3 Hz before 
being cut off. At higher frequencies, the PDS becomes power-law again,
with a slope of roughly -2 (i.e., ``$1/f^2$''). Broad peaks in the range of 
3-9 Hz are present, and might be due to quasi-periodic oscillations. The PDS 
shows interesting spectral dependence: the 1/f component becomes more 
prominent when the low-energy spectrum becomes softer. The difference in
the observed spectral and timing properties between the low and high states 
is qualitatively consistent with a simple ``fluctuating corona'' model. 

\end{abstract}


\section{Introduction}

Cyg~X-1 is a prototype black hole candidate (BHC; see a review by 
\markcite{tanaka1995}Tanaka \& Lewin 1995). It spends most of 
the time in the low (or hard) state where the soft X-ray luminosity 
(2-10 keV) is low and the energy spectrum is characterized by a single 
power-law with a photon index of $\sim 1.5$ (\markcite{liang1984}Liang \& 
Nolan 1984). The X-ray flux varies on all timescales 
down to a few milliseconds, and the power density spectrum (PDS) can be 
characterized by a flat component with a low-frequency cutoff in the range
of $\sim$0.04-0.4 Hz (\markcite{vandk1995}van der Klis 1995 for a review).
An additional steepening of the PDS above 10 Hz was also detected 
(\markcite{belloni1990}Belloni \& Hasinger 1990). Cyg~X-1 has
only occasionally been observed in the high (or soft) state (see reviews by
\markcite{oda1977}Oda 1977 and \markcite{liang1984}Liang \& Nolan 1984), 
and is not well studied in such state. In the high state, the 
power-law photon index varies significantly between 2.6-4.2 below 10 keV 
(\markcite{oda1977}Oda 1977), and mildly between 1.6-2.3 above 10 keV 
(\markcite{liang1984}Liang \& Nolan 1984), while it remains in the range of 
1.3-2.3 (above and below 10 keV) in the low state. Therefore, it seems as 
if the hard power-law component exists in both high and low states, but
the low energy ($<10$ keV) spectrum becomes much softer in the high state. 
No PDS has been reported for Cyg~X-1 in the high state (\markcite{vandk1995}
van der Klis 1995).

In this {\it Letter}, we present the results from the observations of Cyg~X-1 
in its high state by the {\it Rossi X-ray Timing Explorer} (RXTE; 
\markcite{bradt1993}Bradt, Rothschild, \& Swank 1993). 

The ASM (\markcite{levine1996}Levine et al. 1996) light curve
of Cyg~X-1 revealed that it started a transition from the low state to the 
high state on 1996 May 10 (\markcite{cui1996}Cui 1996; Fig.~1a). 
During the transition, the soft X-ray flux (1.3-12 keV) increased by roughly 
a factor of 4, and the ASM hardness ratio, defined as the ratio of the 
3-12 keV count rate to that in the 1.3-3 keV band, shows a steady trend of 
spectral softening (see Fig.~1b). This trend extends to higher energies. 
BATSE observations revealed about a factor of 2
decrease in the 20-100~keV flux (\markcite{zhang1996a}Zhang et al. 
1996a; Fig.~1c). An anti-correlation between the soft (ASM) and hard
(BATSE) X-ray fluxes during the transition, which was observed previously 
(\markcite{dolan1977}Dolan et al. 1977; \markcite{ling1987}Ling et al. 1987), 
was firmly established.

Upon the discovery, a series of public Target-of-Opportunity observations with 
{\it RXTE} were carried out to monitor the temporal and spectral variability 
in the high state. As of 7/15/96, 12 brief pointed observations 
of Cyg~X-1 have been made with {\it RXTE}. Here we concentrate on the 
first 4 observations. Table~1 summarizes the observation time and durations. 

The {\it RXTE} mission is optimized for observing fast X-ray variability 
in a broad energy range. For the first time, $\mu s$ timing resolution is 
achieved for both the PCA and HEXTE instruments 
(\markcite{bradt1993}Bradt, Rothschild, \& Swank 1993), which share a common 
$1^{\circ}$ field-of-view (FWHM). The PCA has a collecting area of about 6500 
$cm^2$ and covers an energy range from 2 to 60 keV with moderate energy 
resolution ($\sim 18\%$ at 6 keV). The HEXTE has a total effective area of 
about 1600 $cm^2$ in two clusters. It covers a wide energy range from 
about 15 to 250 keV with an energy resolution of $\sim 16\%$ at 60 keV.

\section{Spectral Analysis}

We have taken a conservative approach to spectral modelling, 
since the instrument calibration is still preliminary. Due to the high
intensity of Cyg X-1, we have used earth-occultation data to estimate
the total PCA background. This is adequate because the diffuse X-ray 
background contributes less than 1\% of the source 
flux below 20 keV, where systematic uncertainties are dominant, and 
is negligible compared to the instrument background above 20 keV 
(\markcite{jahoda1996}Jahoda et al. 1996). A 2\% systematic uncertainity 
has been added to the PCA data to represent uncertainties in the response 
matrix calibration.

Background-subtraction is straightforward for the HEXTE data because 
the two clusters in HEXTE alternately rock on and off source to provide 
nearly simultaneous background measurement. However, unmeasured deadtime 
effects due to large energy losses from charged particles are significant, so
we allowed the relative normalization between the PCA and HEXTE to vary during
the spectral fitting. We then verified that the resulting relative 
normalizations were consistent with what we expected from the magnitude of the
known deadtime deficit.

The observed X-ray spectrum can be described by a model consisting 
of a soft blackbody component and a broken power-law with a high energy 
cutoff. The best-fit model parameters are listed in Table~2. The 
uncertainties shown represent 90\% confidence intervals. Note that, due to 
the coupling 
between the two spectral components at low energies, we derived the
uncertainty for the blackbody temperature first, froze it, and then
derived the uncertainties for other parameters. The derived $N_{H}$ value is 
about 20 times larger higher than the interstellar value 
($\sim 6.2\times 10^{21} cm^{-2}$) (\markcite{balu1991}Ba{\l}uci\'{n}ska 
\& Hasinger 1991). This could be due to large internal absorption by matter
in the binary system. The stellar wind from the supergiant companion 
may be responsible for providing a large amount of intervening material. 
However, since we only see the tail of the blackbody spectrum in the PCA
energy band, this result can be quite uncertain due to systematic 
uncertainties at the lowest energies. The spectrum was fit again with 
$N_{H}$ fixed at the interstellar value, and the results are also shown
in Table~2 for comparison. The blackbody component is either not needed or 
very insignificant in each of the 4 cases although these fits have
consistently higher $\chi^2$. 

Fortunately, Cyg~X-1 was also observed by {\it ASCA} at about the same time 
as the third {\it RXTE} observation (\markcite{dotani1996}Dotani et al. 
1996). The {\it ASCA} spectrum was modelled with a soft blackbody component 
of temperature $kT=0.34\pm 0.02$ keV and a power-law with a photon index of 
$2.4\pm 0.1$ (\markcite{dotani1996}Dotani et al. 1996). Their best-fit $N_{H}$ 
value is $\simeq 3.2\times 10^{21} cm^{-2}$ (\markcite{negoro1996}Negoro 
1996). These results agree with ours (in the case of low $N_{H}$) reasonably 
well (see Table~2). 

The results in Table~2 show that the low energy X-ray spectrum (i.e., 
$\alpha_1$) varies significantly in the high state on a timescale of days. 
Above the break energy ($\sim$11 keV) the spectral shape changes little. 
As an example, Fig.~2 shows a combined PCA/HEXTE photon spectrum for the 
first observation. 

\section{Timing analysis}

From each of 4 PCA observations we chose a contiguous stretch of data 2048 
(or 4096) seconds long, and generated a PDS in 3 energy bands:
2-6.5 keV, 6.5-13.1 keV, and 13.1-60 keV. The results were then 
logarithmically rebinned to reduce scatter at high frequencies. The 
resulting spectra are shown in Fig.~3. The fractional rms amplitude squared 
is defined as Leahy normalized power (with Poisson noise power subtracted)
divided by the mean source count rate (\markcite{vandk1995}
van der Klis 1995). The PDS's have also been corrected for instrumental 
artifacts due to electronic deadtime and very high energy events 
({\markcite{will1996}Zhang et al. 1996d). The PDS shows roughly
the same shape in different energy bands for a given observation.

In the second observation, the energy spectrum was seen to be the hardest. 
At this time, the PDS can be characterized by a red noise component with a 
characteristic shape of 1/f at low frequencies (less than about 15 mHz), 
followed by a white noise component that extends to about 1 Hz, above which 
it is cut off. At higher frequencies, the PDS becomes power-law again, with 
a slope of roughly -2, i.e., ``$1/f^2$''. A broad peak is detected at about 
3.6 Hz, and might be due to quasi-periodic oscillations.
When the energy spectrum is softer in the first and third observations,
the 1/f noise is more significant, and another feature at around
9 Hz becomes apparent. Therefore, the PDS shows interesting spectral 
dependence: the 1/f component becomes more prominent when the low energy 
spectrum becomes softer. The PDS is eventually dominated by the 1/f noise
in the fourth observation when the energy spectrum is the softest. Similar 
spectral dependence of the PDS was also observed in another BHC, Nova Muscae 
1991 (\markcite{miyam1995}Miyamoto 1995), and maybe common in BHCs. 
In the last observation, the broad peaks at $\sim$3.6 and 9 Hz disappeared,
but a fit to the PDS with a broken power-law reveals another broad feature 
that centers at $\sim$6 Hz.

\section{Discussion}

We interpret the soft blackbody component as the emission from a geometrically
thin, optically thick cool accretion disk. The soft X-ray photons are 
Compton upscattered by a geometrically thick, optically thin corona 
surrounding the disk to produce the observed hard X-ray emission 
(\markcite{liang1984}Liang \& Nolan 1984, and references therein). Then
the spectrum can be approximated by a thermal component around the blackbody 
temperature, a power-law ($\alpha_1$) at energies just above, and a flatter 
power-law component ($\alpha_2$) at still higher energies before being cut 
off beyond $kT_e$, where $T_e$ is the electron temperature of the corona 
(\markcite{liang1984}Liang \& Nolan 1984, and also see discussion by 
\markcite{ebisawa1996}Ebisawa et al. 1996). 
However, the observed high energy cutoff is so gradual that models with a 
single electron temperature (e.g., \markcite{sunyeav1980}Sunyeav \& Titarchuk
1980) fail to fit the high energy portion of the 
spectrum. This slow high energy cutoff can be explained by invoking a
stratified hot electron corona (\markcite{skibo1995}Skibo \& Dermer 1995). 
A similar low-energy spectral shape was observed in the low state by {\it 
ASCA} (\markcite{ebisawa1996}Ebisawa et al. 1996). In their case, the 
soft blackbody component had a lower temperature ($kT\simeq 0.1$ keV), and
the broken power-law was flatter ($\alpha_1=1.92$ and $\alpha_2=1.71$),
with a lower break energy ($\sim 3.4$ keV). 

In the high state the observed PDS shows a distinct 1/f component
at low frequencies, which was not seen in the low state. This component
may be due to the superposition of random accretion ``shots'' with 
long lifetimes (see discussion by \markcite{belloni1990}Belloni \& Hasinger 
1990). Theoretical models have been proposed to associate these shots with 
instabilities in the accretion disk (e.g., \markcite{minesh1994}Mineshige et 
al. 1994). Perhaps, the 1/f noise increases its power when the mass 
accretion rate is higher, which would explain why it was not observed in the 
low state (presumably with a lower accretion rate). Its dominance in
the fourth observation seems to support this, although there is 
indication that the bolometric luminosity changes little going from the low
state to the high state (\markcite{zhang1996c}Zhang et al. 1996c). 
The fourth observation may mark the start of the ``true'' high state, 
which is characterized by the soft energy spectrum and dominant 1/f noise, 
following a ``settling period''. This is supported by the results from 
subsequent observations (\markcite{cuietal1996}Cui, Focke, \& Swank 1996). 
Similar
power-law PDS's were observed in soft X-ray transient BHCs, Nova Muscae 1991 
(\markcite{miyam1995}Miyamoto 1995) and Nova Sco 1994 (\markcite{zhang1996b}
Zhang et al. 1996b), in their outburst states, and may be common among BHCs 
in their high states.

The white noise (or flat) component has been seen in both high and low 
states. It may be due to statistical fluctuations in the mass accretion 
stream near the inner edge of the accretion disk where dynamical 
timescale is very short (compared to the frequency range that has been 
covered), similar to the ``shot noise'' in many electronic systems 
(\markcite{ziel1986}van der Ziel 1986). The hot corona in the system can 
act as a low-pass filter that cuts off the 
white noise at some characteristic frequency to produce the observed 
``flat-top'' PDS shape. In this model, the cutoff frequency is determined
by the characteristic photon escape time. Then, the higher cutoff frequency
observed in the high state ($\sim$1 Hz, compared to $\sim$0.1 Hz in the
low state) can be explained by a smaller corona due to more efficient
local cooling provided by a higher mass accretion rate. Because the corona 
is smaller 
in the high state, the number of scatterings that an X-ray photon 
experiences is, on average, less, therefore the emerging hard X-ray 
spectrum is softer, which agrees qualitatively with the observations. 
In the fourth observation, the white noise becomes
negligible compared to the 1/f noise, and the observed PDS shape 
is consistent with the 1/f noise being cutoff at around 16 Hz (thus
becoming $1/f^2$ at higher frequencies). Therefore the corona seems to be even 
smaller in this case, which would explain the softest energy 
spectrum among all observations (see Table~2). 

What is the origin of the broad PDS peaks detected between 3-9 Hz? Similar
features are often seen on the noise spectra of solid state devices.
They are thought to be produced by a charge generation and recombination 
process due to the existence of impurities in semi-conductor material 
(\markcite{ziel1986}van der Ziel 1986). In this case, the noise spectra are 
characterized by a flat component below a characteristic frequency and a 
power-law shape of $1/f^2$ above, which is the same as that of exponential 
accretion ``shots'' (\markcite{belloni1990}Belloni \& Hasinger 1990). 
Similar processes might be involved 
in mass accretion due to shocks and turbulences in the accretion disk that 
disrupt the flow. As a result, the lifetime distribution of the 
accretion shots may not be random, but limited to certain values due to 
activation or some other critical conditions (e.g., \markcite{minesh1994}
Mineshige et al. 1994) in such processes. It seems more likely, however, that 
the these features are related to resonances in the fluctuating corona 
because they appears to be more prominent at higher energies.

{\it Note added in manuscript}: after we submitted this paper, we became aware
of another paper submitted at about the same time by \markcite{belloni1996}
Belloni et al. (1996) which is based on some of the same data presented here. 
We have independently reached at some similar conclusions. For 
example, their proposed ``intermediate state'' is similar to our ``settling 
period''.

\acknowledgments
We wish to thank every member of the {\it RXTE} team for the success of the 
mission. We would like to thank J.~Swank and H.~Bradt 
for valuable comments, and E.~H.~Morgan for discussions on timing analysis. 
We are also grateful to an anonymous referee for comments that resulted in 
an improved manuscript. This work is supported in part by NASA Contracts 
NAS5-30612 and NAS5-30720.

\clearpage

\begin{deluxetable}{lccc}
\tablecolumns{4}
\tablewidth{0pc}
\tablecaption{{\it RXTE} Observations of Cyg~X-1}
\tablehead{
 & & \multicolumn{2}{c}{Live Time (s)} \\
\cline{3-4}
\colhead{No.} & \colhead{Observation Time (UT)} & \colhead{PCA} & \colhead{HEXTE\tablenotemark{1}}
}
\startdata
1 & 5/22/96 17:44:00-19:48:00 & 4208 & 1312 \nl
2 & 5/23/96 14:13:00-18:07:00 & 7936 & 5839 \nl
3 & 5/30/96 07:46:00-08:44:00 & 2384 & 2113 \nl
4 & 6/4/96 20:21:00-21:42:00 & 3280 & 2415 \nl
\tablenotetext{1}{Both clusters are included.}
\enddata
\end{deluxetable}

\begin{deluxetable}{lcccccccccc}
\scriptsize
\tablecolumns{11}
\tablewidth{0pc}
\tablecaption{Results of Spectral Analysis}
\tablehead{
 & & blackbody & \multicolumn{3}{c}{broken power-law\tablenotemark{1}} & \multicolumn{2}{c}{high-energy cutoff\tablenotemark{2}} \\
\cline{3-3} \cline{4-6} \cline{7-8} \\
\colhead{No.} & \colhead{$N_H$\tablenotemark{3}}& \colhead{$kT_b$} & \colhead{$\alpha_1$}& \colhead{$\alpha_2$} & \colhead{$E_b$} & \colhead{$E_c$} & \colhead{$E_f$} & \colhead{$\chi^2_{\nu}/dof$} & \colhead{Flux\tablenotemark{4}} & f\tablenotemark{5} \\
 & ($10^{22}\mbox{ }cm^{-2}$) & (keV) & & & (keV) & (keV) & (keV) & & &\%  } 
\startdata
1 & $11.3^{+1.5}_{-1.8}$ & $0.27^{+0.02}_{-0.02}$ & $2.95^{+0.04}_{-0.05}$ & $1.95^{+0.03}_{-0.04}$ & $10.8^{+0.3}_{-0.2}$ & $24^{+4}_{-4}$ & $184^{+28}_{-24}$ & $1.00/176$ & $1.88$ & $35$ \nl
  & $0.62$ (fixed) & \nodata & $2.59^{+0.02}_{-0.01}$ & $1.86^{+0.04}_{-0.04}$ & $10.8^{+0.3}_{-0.3}$ & $21^{+3}_{-18}$ & $144^{+19}_{-18}$ & $1.15/179$ & $1.97$ & \nodata \nl
2 & $10.0^{+2.0}_{-2.1}$ & $0.30^{+0.03}_{-0.02}$ & $2.60^{+0.05}_{-0.05}$ & $1.84^{+0.02}_{-0.02}$ & $11.2^{+0.4}_{-0.3}$ & $24^{+2}_{-2}$ & $154^{+11}_{-11}$ & $1.03/176$ & $1.72$ & $24$ \nl
  & $0.62$ (fixed) & \nodata & $2.20^{+0.01}_{-0.02}$ & $1.79^{+0.03}_{-0.02}$ & $12.0^{+0.04}_{-0.05}$ & $23^{+2}_{-1}$ & $137^{+9}_{-8}$ & $1.55/179$ & $1.80$ & \nodata \nl
3 & $10.9^{+1.5}_{-1.6}$ & $0.27^{+0.02}_{-0.02}$ & $2.98^{+0.05}_{-0.04}$ & $1.91^{+0.03}_{-0.04}$ & $10.8^{+0.2}_{-0.3}$ & $25^{+4}_{-4}$ & $179^{+33}_{-25}$ & $0.83/176$ & $1.38$ & $38$ \nl
  & $0.62$ (fixed) & $0.30^{+0.11}_{-0.09}$ & $2.65^{+0.03}_{-0.03}$ & $1.83^{+0.04}_{-0.05}$ & $10.7^{+0.03}_{-0.03}$ & $22^{+4}_{-4}$ & $141^{+21}_{-18}$ & $0.99/177$ & $1.47$ & $5$ \nl
4 & $14.0^{+1.2}_{-1.3}$ & $0.24^{+0.02}_{-0.01}$ & $3.26^{+0.04}_{-0.04}$ & $2.09^{+0.03}_{-0.04}$ & $10.8^{+0.2}_{-0.2}$ & $24^{+4}_{-4}$ & $216^{+44}_{-32}$ & $0.95/176$ & $1.52$ & $46$ \nl
  & $0.62$ (fixed) & $0.31^{+0.06}_{-0.07}$ & $2.82^{+0.03}_{-0.03}$ & $1.99^{++0.04}_{-0.04}$ & $10.8^{+0.2}_{-0.3}$ & $21^{+2}_{-4}$ & $154^{+22}_{-22}$ & $1.28/177$ & $1.67$ & $9$ \nl
\tablenotetext{1}{$\alpha_1$ and $\alpha_2$ are soft and hard power-law photon indices, respectively, and $E_b$ is the break energy.}
\tablenotetext{2}{$E_c$ is the cutoff energy, and $E_f$ the e-folding energy.}
\tablenotetext{3}{H I column density along the line-of-sight.}
\tablenotetext{4}{The observed 2-10 keV flux (in units of $10^{-8}\mbox{ }erg\mbox{ }cm^{-2}\mbox{ }s^{-1}$).} 
\tablenotetext{5}{The observed fractional 2-10 keV flux from the blackbody component.} 
\enddata
\end{deluxetable}

\newpage
\begin{figure}[t]
\epsfxsize=350pt \epsfbox{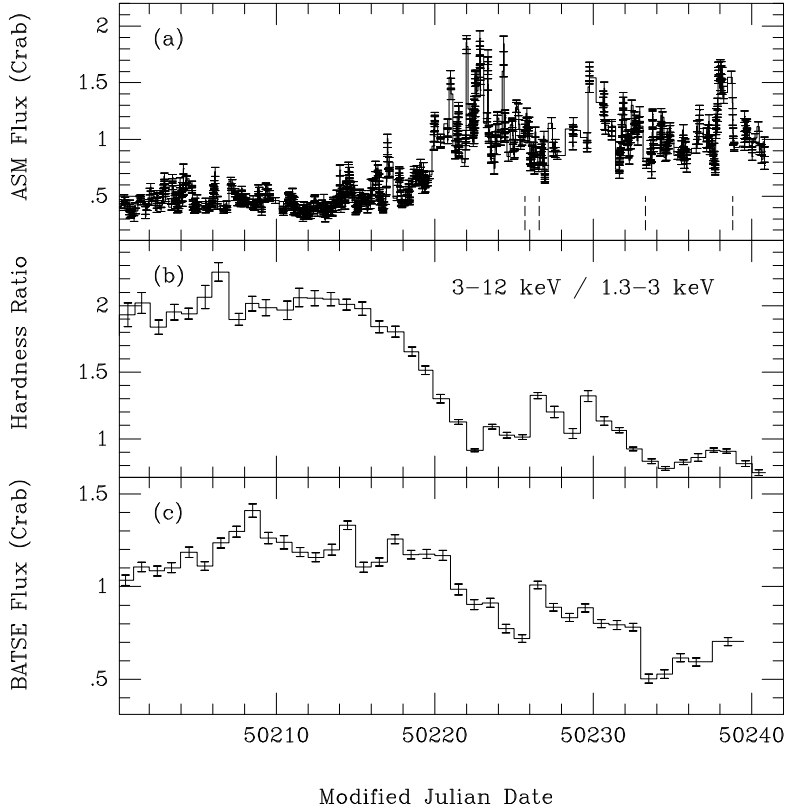}
\caption{(a) The ASM light curve of Cyg~X-1. It comprises 
measurements from individual ``dwells'' with 90-second exposure time.
The vertical dashed-lines indicate when the {\it RXTE} observations were
made. (b) The daily-averaged time series of the ASM hardness ratio 
(3-12 keV/1.3-3 keV); and (c) The daily-averaged BATSE (20-100 keV) 
light curve. MJD 50213.0 corresponds to 1996 May 10 0 h UT.}
\end{figure}

\begin{figure}[t]
\epsfxsize=600pt \epsfbox{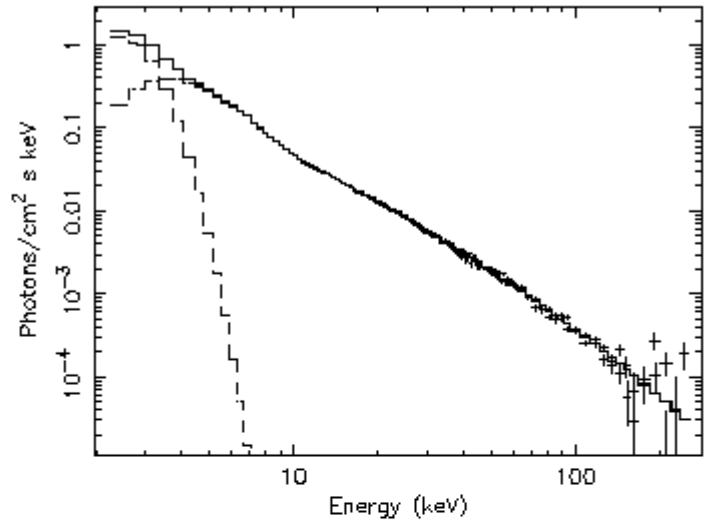}
\caption{The combined PCA/HEXTE photon spectrum of Cyg~X-1 for 
the first observation. For clarity we show only the Poisson error. 
The best-fit model is shown in solid-line, along with each spectral 
component in dashed-lines to show their relative contribution. }
\end{figure}

\begin{figure}[t]
\epsfxsize=350pt \epsfbox{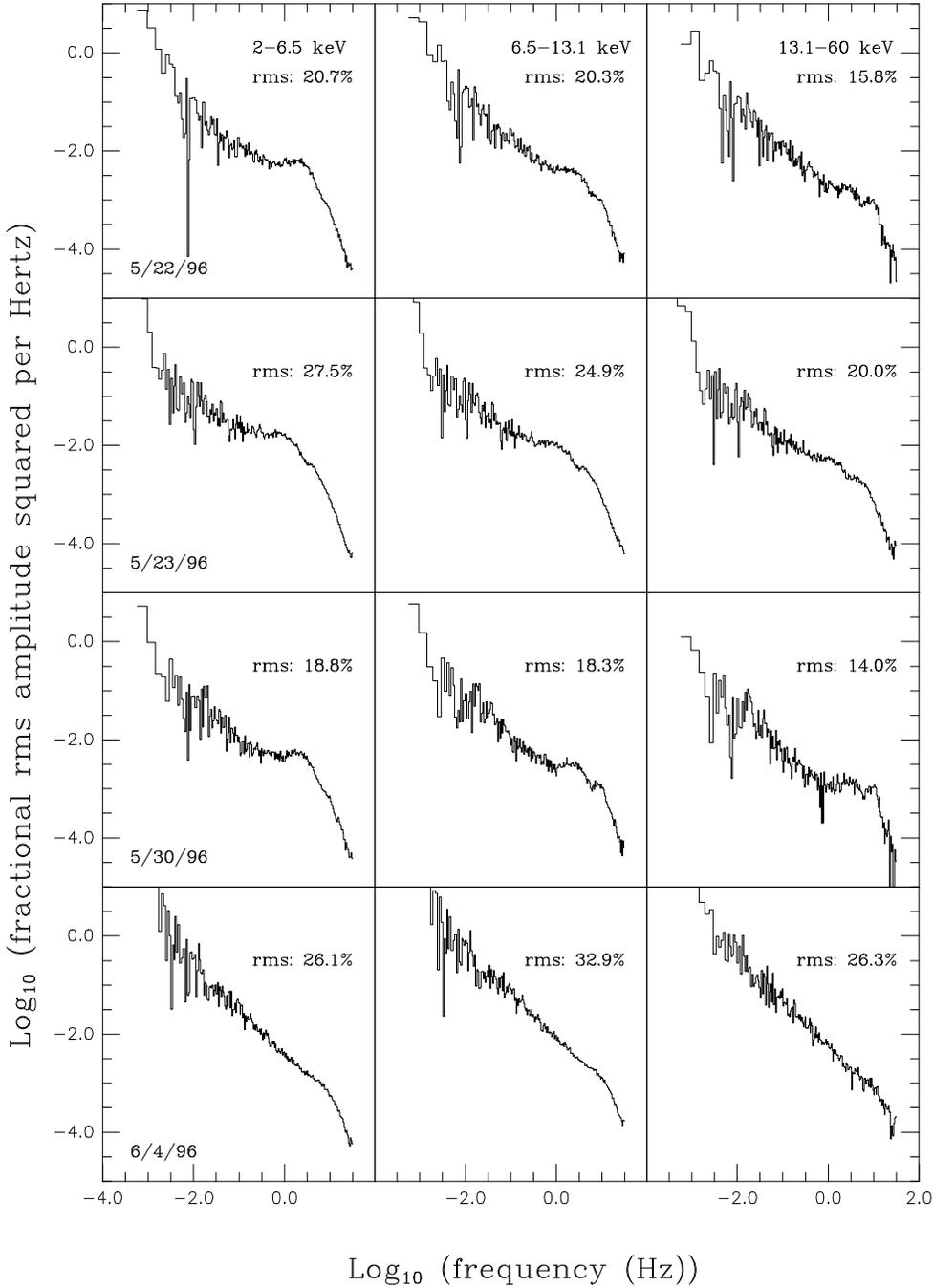}
\caption{The power density spectra in 3 energy bands derived 
from the PCA data. The results from different 
observations are presented in different columns, and each row contains a 
single energy band. Note that the integrated fractional rms noise shown is 
derived in the frequency range of 0.488 (or 0.244) mHz-32 Hz, depending on
the observation.}
\end{figure}

\end{document}